\documentclass[onecolumn,preprintnumbers,amssymb,amsmath,superscriptaddress,letterpaper,nofootinbib]{revtex4}[12pt]

\usepackage{graphicx}
\usepackage{dcolumn}
\usepackage{bm}
\usepackage{bbold}
\usepackage{natbib}
\usepackage{pstricks}
\usepackage{epsfig}
\usepackage{epstopdf}
\usepackage{amsmath}
\usepackage{tikz,lmodern,amssymb}
\usetikzlibrary{calc}
\usepackage{mathrsfs}
\usepackage[toc]{appendix}
\usepackage{booktabs}
\usepackage[colorlinks,linkcolor=magenta,anchorcolor=blue,citecolor=green]{hyperref}

\newcommand{\be}{\begin{equation}}
	\newcommand{\ee}{\end{equation}}
\newcommand{\bea}{\begin{eqnarray}}
	\newcommand{\eea}{\end{eqnarray}}

\begin{document}

	\title{Bimodal mass primordial black holes in three-field inflation}
	
	\author{M. M. Nasrolahzadeh}
	\email{m.nasrolahzadeh@mail.sbu.ac.ir}
	\affiliation{Department of Physics, Shahid Beheshti University, G.C., Evin, Tehran 19839, Iran}
	
	\author{N. Khosravi}
	\email{nima@sharif.edu}
	\affiliation{Department of Physics, Sharif University of Technology, Tehran 11155-9161, Iran}
	
	\date{\today}

	\begin{abstract}
		In this paper, we study the evolution of curvature and entropy perturbations in three-field inflation with non-canonical kinetic terms. Using the kinematic basis in field space, we characterize the background trajectory through its turning rate and non-planar torsion. We also emphasize the inherited structure of perturbations from the kinematic system, which detaches the distinct roles of turning rate and torsion in the interactions among perturbations. We present and study numerically a three-phase model with an intermediate decelerated expansion stage in which two separated transient epochs amplify two distinct families of modes, generating a double-peaked curvature power spectrum. This double-peaked power spectrum leads to the formation of primordial black holes with two characteristic masses, $M_1\sim2\times10^{-12}M_\odot$ and $M_2\sim52M_\odot$ explaining, simultaneously, both dark matter and the LIGO-Virgo-KAGRA binary black hole mass range. 
	\end{abstract}
	
	\maketitle

	\section{Introduction}
	Inflation \cite{Linde:1981mu,Guth:1980zm,Starobinsky:1982ee} as the leading theory of the early universe in addition to resolving the fine-tuning problems of the Big-Bang model, also explains the origin of temperature fluctuations in the cosmic microwave background (CMB) and the large-scale structures. It predicts nearly scale-invariant and Gaussian statistics for the initial fluctuations, which are compatible with the CMB observation with high accuracy \cite{Planck:2018jri}.
	
	The inflationary scenario as an initial condition for our universe is often modeled with a single scalar field. However, models derived from high energy physics (e.g. string theory, supergravity) involve multiple scalar fields living in an abstract curved manifold $\mathcal{M}_\phi$ \cite{Berglund:2009uf,Achucarro:2018vey}. The possible nontrivial geometry of this manifold forces the background solution to follow a non-geodesic trajectory, whose important information can be characterized by quantities such as turning rate, torsion\footnote{As we will discuss more later, this torsion is the twist rate of the background trajectory, not the geometric torsion of manifold $\mathcal{M}_\phi$.} and higher-order bending parameters. Despite single field inflation, which only generates adiabatic (curvature) perturbations, multifield models with $\mathcal{N}$ scalar fields possess $\mathcal{N}-1$ entropy (isocurvature) modes, whose dynamics can modify the observables \cite{Cespedes:2012hu,Achucarro:2016fby}. Since the CMB (as the best probe of inflation) is a localized event in both time and spatial manner, the small value of entropy perturbations, consistent with the CMB results \cite{Planck:2018jri}, does not exclude their presence during the inflationary era. Inflation is usually regarded as a connected period of accelerated expansion. However, in multi-phase inflationary scenarios, depending on the strength of the phase transitions, inflation can have a break with an intermediate decelerated expansion era. The occurrence of such a break can imprint on the observables as a step-like spectrum in curvature \cite{Polarski:1992dq,Polarski:1994rz}. 
	
	On the other hand, primordial black holes (PBHs) \cite{Zeldovich:1967lct,Hawking:1971ei,Carr:1974nx} are one of the most plausible dark matter candidates, which can form from the gravitational collapse of overdensities in the early universe. For this to happen, the power spectrum of curvature perturbations $\mathcal{R}$ must be amplified seven orders of magnitude on small scales compared to its value measured in the CMB. One can do this by adding a feature to the inflaton potential \cite{Mishra:2019pzq,Atal:2019cdz,Ragavendra:2021qdu} or a sharp turn in the dynamics resulting from multifield models \cite{Palma:2020ejf,Fumagalli:2020adf}, etc. Because of the correspondence between PBHs mass and the wavelength of perturbations, PBHs can be a suitable probe of the curvature perturbations at small scales. PBHs may also have the potential to explain the gravitational waves (GWs) in the mass range of black hole binaries detected by the LIGO-Virgo-KAGRA collaborations \cite{LIGOScientific:2025slb}. This motivates the search for inflationary models with two distinct peaks in the curvature power spectrum, which would lead to the formation of PBHs with two characteristic mass scales, addressing dark matter and GW signals simultaneously \cite{Inomata:2018cht,Ballesteros:2018wlw}.
	
	In this work, we explore three-field inflation, the simplest setup with multiple entropy modes, as a good framework to investigate whether different entropy modes play distinct roles. We employ the Frenet–Serret system to define an orthogonal basis along the background trajectory in a curved field space and decompose the perturbations into adiabatic and entropy components. Through the numerical computations, we study the effect of turns and break in three-field inflation on the PBHs mass spectrum. 
	
	Though we are focusing on three-field inflation but we see this proposal in a more general multifield scenario. At the first look, it is natural to ask why one should think about a multifield scenario if the simpler single field inflation can (fully) describe the observations? To justify why we need to think about more complex scenarios (e.g. for the inflation) we need to look more carefully at the standard model of cosmology, $\Lambda$CDM. This model, which is very good in agreement with the observations is the most simple scenario which could be proposed. The gravitational field is described by the Einstein general relativity, which is the simplest scenario for the gravity\footnote{By requesting the universal and attractive gravitational field, the simplest scenario is the massless spin-2 mediator which is given uniquely by the Einstein gravity.}. To address the missing mass, the simplest proposal is a matter field with no interaction which is the CDM. The initial conditions are set by the single field inflation in (almost) its simplest form. Finally, to address the late time acceleration, the simplest scenario is the cosmological constant i.e. $\Lambda$. This means it seems the universe is very fine-tuned to be describable by the simplest theoretical model. A viewpoint which can be seen as a solution to this kind of fine-tuning is to think the $\Lambda$CDM naturally emerges from a very complex model e.g. see \cite{Khosravi:2020lbf}. One of our motivations for the current work is exactly the above question, and we will see that assuming a more complicated inflationary model will have some observable consequences.

	\section{General three-field inflation}
	For the case of three scalar fields $\phi^a=(\phi,\chi,\psi)$ with minimal coupling, the most general action in four dimensional curved spacetime is
	\begin{equation}\label{eq:FullAction}
		S=\int d^4x\sqrt{-g}\left[\frac{M^2_P}{2}R-\frac{1}{2}\mathcal{G}_{ab}g^{\mu\nu}\partial_\mu\phi^a\partial_\nu\phi^b-V(\phi^a)\right],
	\end{equation}
	where $M_P$ is the reduced Planck mass, $M_P\equiv1/\sqrt{8\pi G}$. These three fields span a three-dimensional (curved) field space $\mathcal{M}_\phi$, whose geometry is constructed by the metric $\mathcal{G}_{ab}$. In this work we consider this diagonal form for the  metric
	\begin{align}\label{eq:fsmetric}
		\mathcal{G}_{ab}\equiv
		\begin{pmatrix}
			1&0&0\\
			0&e^{2b(\phi)}&0\\
			0&0&e^{2y(\phi)+2p(\chi)}
		\end{pmatrix}.
	\end{align}
	When $b(\phi)=y(\phi)=p(\chi)=0$, the action
	(\ref{eq:FullAction}) returns to its canonical form. The dependence on $\phi$ and $\chi$ in the kinetic energy of the third field can arise in the 4D effective action obtained from heterotic M-theory \cite{Copeland:2001zp}. The two-field models based on the leading $2\times2$ principal block of the field space metric (\ref{eq:fsmetric}) have been extensively studied in the literature \cite{DiMarco:2002eb,Lalak:2007vi,Cremonini:2010ua,Avgoustidis:2011em,Braglia:2020fms,DeAngelis:2023fdu}. 
	\subsection{Background evolution}
	At the homogeneous level, we use the spatially flat FRW metric
	\begin{equation}
		ds^2=-dt^2+a^2(t)\delta_{ij}dx^idx^j.
	\end{equation}
	The equations of motion regarding the nontrivial field space metric in FRW spacetime are
	\begin{equation}\label{eq:BackgroundEqns}
		\begin{split}
			&\ddot{\phi}+3H\dot{\phi}-b_\phi e^{2b}\dot{\chi}^2-y_\phi e^{2y+2p}\dot{\psi}^2+V_\phi=0,\\
			&\ddot{\chi}+\left(3H+2b_\phi\dot{\phi}\right)\dot{\chi}-p_\chi e^{2y+2p-2b}\dot{\psi}^2+e^{-2b}V_\chi=0,\\
			&\ddot{\psi}+\left(3H+2y_\phi\dot{\phi}+2p_\chi\dot{\chi}\right)\dot{\psi}+e^{-2y-2p}V_\psi=0,\\
			&H^2=\frac{1}{3M^2_P}\left(\frac{1}{2}\dot{\phi}^2+\frac{1}{2}e^{2b}\dot{\chi}^2+\frac{1}{2}e^{2y+2p}\dot{\psi}^2+V\right),\\
			&\dot{H}=-\frac{1}{2M^2_P}\left(\dot{\phi}^2+e^{2b}\dot{\chi}^2+e^{2y+2p}\dot{\psi}^2\right),
		\end{split}
	\end{equation}
	where $H$ is the Hubble rate, $H\equiv\dot{a}/a$, and a dot denotes a derivative with respect to cosmic time $t$. Subscripts denote a derivative with respect to the corresponding field. All the equations in (\ref{eq:BackgroundEqns}) are not independent and the fourth one (the Friedmann equation) can be used as a constraint. Also it is useful to introduce the first slow-roll parameter $\epsilon\equiv-\dot{H}/H^2$ and the number of e-folds $N_e\equiv\int Hdt$.
	
	The background trajectory in three-dimensional field space has one tangent vector and two preferred normal vectors perpendicular to it. The tangent vector is the unit velocity vector
	\begin{equation}
		\mathscr{T}^a\equiv\left(\frac{\dot{\phi}}{\dot{\sigma}},\frac{\dot{\chi}}{\dot{\sigma}},\frac{\dot{\psi}}{\dot{\sigma}}\right),
	\end{equation}
	where $\dot{\sigma}\equiv\sqrt{\mathcal{G}_{ab}\dot{\phi}^a\dot{\phi}^b}$ is the rate of change in the field vacuum expectation value. The turning rate of the trajectory can be computed from the covariant derivative of this vector 
	\begin{equation}
		\Omega\equiv\sqrt{\mathcal{G}_{ab}\mathcal{D}_t\mathscr{T}^a\mathcal{D}_t\mathscr{T}^b},
	\end{equation}
	where $\mathcal{D}_t\mathscr{T}^a=\dot{\mathscr{T}}^a+\mathbb{\Gamma}^a_{bc}\mathscr{T}^c\dot{\phi}^b$, and $\mathbb{\Gamma}^a_{bc}$ is the connection associated to the metric $\mathcal{G}_{ab}$.
	There is freedom to choose any two vectors orthogonal to 
	$\mathscr{T}$ as the directions of the normal vectors. Here we use the kinematic Frenet-Serret system to determine these two vectors \cite{kreyszig2013differential}. The first normal vector is the normalized covariant derivative of the tangent vector
	\begin{equation}
		\mathscr{N}^a\equiv\frac{1}{\Omega}\mathcal{D}_t\mathscr{T}^a.
	\end{equation}
	By using exterior algebra, one can obtain the binormal vector as
	\begin{equation}
		\mathscr{B}^a\equiv\frac{1}{\sqrt{\mathrm{det}\mathcal{G}}}\varepsilon^{abc}\mathscr{T}_b\mathscr{N}_c,
	\end{equation}
	where we use this convention for the Levi-Civita coefficients $\varepsilon^{\phi\chi\psi}=1$. Also the quantity that measures the twisting of the $\mathscr{T}-\mathscr{N}$ osculating plane is the torsion
	\begin{equation}
		\Xi\equiv-\mathscr{N}_a\mathcal{D}_t\mathscr{B}^a.
	\end{equation}
	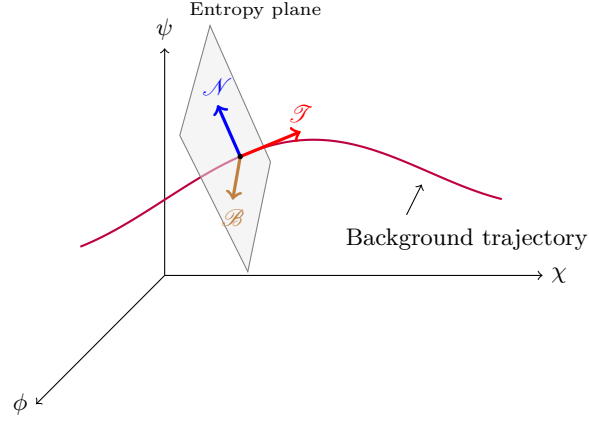
\begin{figure}
		\begin{tikzpicture}
			\draw[->] (0,0) -- (5,0) node[right] {$\chi$};
			\draw[->] (0,0) -- (0,3) node[above] {$\psi$};
			\draw[->] (0,0) -- (-1.7,-1.7) node[left] {$\phi$};
			\draw[thick, purple, domain=-1:4, samples=100]
			plot (
			{\x},
			{0.6*sin(\x r) + 1},
			{-0.3*\x}
			);
			\coordinate (O) at (1,1.57);
			\filldraw[
			fill=gray!15,
			draw=gray,
			thin,
			fill opacity=0.5
			]
			(0.6,3.3) 
			-- (1.4,1.5) 
			-- (1.1,0.05) 
			-- (0.2,1.85) 
			-- cycle;
			\node[black,font=\scriptsize] at (1.2, 3.5) {Entropy plane};
			\draw[->, red, very thick] (O) -- (1.8,1.9) node[above] {$\mathscr{T}$};
			\draw[->, blue, very thick] (O) -- (0.7,2.25) node[above] {$\mathscr{N}$};
			\draw[->, brown, very thick] (O) -- (0.9,1) node[below] {$\mathscr{B}$};
			\node[circle, fill=black, inner sep=0.7pt] at (O) {}; 
			\draw[->, black, thin] (3.2,0.8) -- (3.4,1.2);
			\node[black,font=\small] at (4, 0.48) {Background trajectory};
		\end{tikzpicture}
		\caption{A schematic illustration of the triad of tangent and normal vectors to the background trajectory.}
		\label{fig:TNB}
	\end{figure}
	In summary, we can present this matrix form of the Frenet-Serret system\footnote{Note that the Frenet-Serret system becomes ill-defined when $\Omega=0$.}
	\begin{align}
		\mathcal{D}_t
		\begin{pmatrix}
			\mathscr{T}^a\\
			\mathscr{N}^a\\
			\mathscr{B}^a
		\end{pmatrix}
		=
		\begin{pmatrix}
			0&\Omega&0\\
			-\Omega&0&\Xi\\
			0&-\Xi&0
		\end{pmatrix}
		\begin{pmatrix}
			\mathscr{T}^a\\
			\mathscr{N}^a\\
			\mathscr{B}^a
		\end{pmatrix}.
	\end{align}
	An illustration of these three vectors, which will determine the directions of the adiabatic and entropy perturbations, is shown in FIG. \ref{fig:TNB}. The important consequence of this kinematic basis is 
	\begin{align}
		V^a=-\left(\ddot{\sigma}+3H\dot{\sigma}\right)\mathscr{T}^a-\Omega\dot{\sigma}\mathscr{N}^a,
	\end{align}
	which states that the potential gradient has no component along the binormal direction, i.e. $V_a\mathscr{B}^a=0$.
	\subsection{Linear perturbations}
	Because of the absence of anisotropic stress tensor in the scalar fields system at linear order, we consider this form of the perturbed spacetime metric in the Newtonian gauge
	\begin{equation}
		ds^2=-(1+2\Phi)dt^2+a^2(t)(1-2\Phi)\delta_{ij}dx^idx^j.
	\end{equation}
	We also decompose the scalar fields into background and perturbation parts, where the background evolution is governed by (\ref{eq:BackgroundEqns}),
	\begin{equation}
		\phi(t,\mathbf{x})=\phi(t)+\delta\phi(t,\mathbf{x}),\quad\chi(t,\mathbf{x})=\chi(t)+\delta\chi(t,\mathbf{x})\quad \text{and}\quad\psi(t,\mathbf{x})=\psi(t)+\delta\psi(t,\mathbf{x}).
	\end{equation}
	The Klein-Gordon equations in the Fourier space\footnote{Note that we drop the subscript $\mathbf{k}$ from the notation of perturbations in the Fourier space.} are given by
	\begin{align}\label{eq:KleinGordon1}
		\ddot{\delta\phi}+3H\dot{\delta\phi}&+\left[\frac{k^2}{a^2}+V_{\phi\phi}-\left(2b^2_\phi+b_{\phi\phi}\right)e^{2b}\dot{\chi}^2-\left(2y^2_\phi+y_{\phi\phi}\right)e^{2y+2p}\dot{\psi}^2\right]\delta\phi\notag\\
		&+\left(V_{\phi\chi}-2y_\phi p_\chi e^{2y+2p}\dot{\psi}^2\right)\delta\chi-2b_\phi e^{2b}\dot{\chi}\dot{\delta\chi}+V_{\phi\psi}\delta\psi-2y_\phi e^{2y+2p}\dot{\psi}\dot{\delta\psi}=4\dot{\phi}\dot{\Phi}-2\Phi V_\phi,
	\end{align}
	\begin{align}\label{eq:KleinGordon2}
		\ddot{\delta\chi}&+\left(3H+2b_\phi\dot{\phi}\right)\dot{\delta\chi}+\left[\frac{k^2}{a^2}+e^{-2b}V_{\chi\chi}-\left(2p^2_\chi+p_{\chi\chi}\right)e^{2y+2p-2b}\dot{\psi}^2\right]\delta\chi\notag\\
		&+\left[2b_{\phi\phi}\dot{\phi}\dot{\chi}+e^{-2b}V_{\phi\chi}-2b_\phi e^{-2b}V_\chi+2(b_\phi- y_\phi )p_\chi e^{2y+2p-2b}\dot{\psi}^2\right]\delta\phi\notag\\
		&+2b_\phi\dot{\chi}\dot{\delta\phi}
		-2p_\chi e^{2y+2p-2b}\dot{\psi}\dot{\delta\psi}+e^{-2b}V_{\chi\psi}\delta\psi\notag\\
		&=4\dot{\chi}\dot{\Phi}-2e^{-2b}V_\chi\Phi,
	\end{align}
	and
	\begin{align}\label{eq:KleinGordon3}
		\ddot{\delta\psi}&+\left(3H+2y_\phi\dot{\phi}+2p_\chi\dot{\chi}\right)\dot{\delta\psi}+\left[\frac{k^2}{a^2}+e^{-2y-2p}V_{\psi\psi}\right]\delta\psi\notag\\
		&+\left[2y_{\phi\phi}\dot{\phi}\dot{\psi}-2y_\phi e^{-2y-2p}V_\psi+e^{-2y-2p}V_{\psi\phi}\right]\delta\phi+\left[2p_{\chi\chi}\dot{\chi}\dot{\psi}-2p_\chi e^{-2y-2p}V_\psi+e^{-2y-2p}V_{\psi\chi}\right]\delta\chi\notag\\
		&+2y_\phi\dot{\psi}\dot{\delta\phi}+2p_\chi\dot{\psi}\dot{\delta\chi}=4\dot{\psi}\dot{\Phi}-2e^{-2y-2p}V_\psi\Phi.
	\end{align}
	The perturbed Einstein equations as the energy-momentum constraints are, respectively,
	\begin{align}
		3H\left(\dot{\Phi}+H\Phi\right)&+\dot{H}\Phi+\frac{k^2}{a^2}\Phi\notag\\
		&\hspace{-1.18cm}=-\frac{1}{2M^2_P}\left(\dot{\phi}\dot{\delta\phi}+e^{2b}\dot{\chi}\dot{\delta\chi}+e^{2y+2p}\dot{\psi}\dot{\delta\psi}+\left(b_\phi e^{2b}\dot{\chi}^2+y_\phi e^{2y+2p}\dot{\psi}^2\right)\delta\phi+p_\chi e^{2y+2p}\dot{\psi}^2\delta\chi+V_\phi\delta\phi+V_\chi\delta\chi+V_\psi\delta\psi\right),\label{eq:G00}\\
		&\hspace{-2.1cm}\dot{\Phi}+H\Phi=\frac{1}{2M^2_P}\left(\dot{\phi}\delta\phi+e^{2b}\dot{\chi}\delta\chi+e^{2y+2p}\dot{\psi}\delta\psi\right).\label{eq:G0i}
	\end{align}
	Although the Bardeen potential $\Phi$ can be determined algebraically from equations (\ref{eq:G00}) and (\ref{eq:G0i}), its evolution can also be obtained from the spatial part of the Einstein equations
	\begin{align}
		\ddot{\Phi}&+4H\dot{\Phi}+\left(3H^2+\dot{H}\right)\Phi\notag\\
		&=\frac{1}{2M^2_P}\left(\dot{\phi}\dot{\delta\phi}+e^{2b}\dot{\chi}\dot{\delta\chi}+e^{2y+2p}\dot{\psi}\dot{\delta\psi}+\left(b_\phi e^{2b}\dot{\chi}^2+y_\phi e^{2y+2p}\dot{\psi}^2\right)\delta\phi+p_\chi e^{2y+2p}\dot{\psi}^2\delta\chi-V_\phi\delta\phi-V_\chi\delta\chi-V_\psi\delta\psi\right).
	\end{align}
	
	The initial conditions for the field perturbations at some initial conformal time $\tau$, when the mode $k$ is sufficiently subhorizon, are given by the Bunch-Davies vacuum\footnote{The Bunch-Davies vacuum $\left(e^{-ik\tau}/\sqrt{2k}\right)$ actually is associated to the canonically normalized fields $u_I\equiv a\sqrt{\mathcal{G}_{II}}\delta\phi^I$.}
	\begin{equation}
		\delta\phi(\tau)=\frac{e^{-ik\tau}}{a(\tau)\sqrt{2k}},\quad \delta\chi(\tau)=\frac{e^{-ik\tau}}{a(\tau)e^{b}\sqrt{2k}},\quad \delta\psi(\tau)=\frac{e^{-ik\tau}}{a(\tau)e^{y+p}\sqrt{2k}}.
	\end{equation}
	With these in hand, one can find the initial condition for the Bardeen potential using (\ref{eq:G00}) and (\ref{eq:G0i}). In the subhorizon limit, curvature and entropy perturbations are decoupled. Therefore, to preserve this statistical independence, we integrate the equations of motion three times, each time imposing the Bunch-Davies initial condition on one of the perturbations, following the procedure of 
	\cite{Tsujikawa:2002qx}
	in the two-field and
	\cite{Ashoorioon:2025iid}
	in the three-field models\footnote{It can be shown that this approach at the linear level is equivalent to the in-in formalism \cite{Chen:2015dga}.}.
	
	Unlike the adiabatic perturbation, which always has a unique direction, entropy perturbations belong to an $(\mathcal{N}-1)$ -dimensional hypersurface orthogonal to the adiabatic direction, and this exhibits a freedom to choose between all compatible independent entropy bases, physical or unphysical. One way to fix this freedom is to use the kinematic system we introduced earlier \cite{Kaiser:2012ak,Cespedes:2013rda,Christodoulidis:2022vww,Christodoulidis:2023eiw,Basiouris:2025yir}. The perturbations along the adiabatic and entropy directions are obtained by projecting the Mukhanov-Sasaki gauge-invariant variables onto $\mathscr{TNB}$ vectors
	\begin{equation}\label{eq:gaugeperturbations}
		Q_\sigma=\mathscr{T}_IQ^I,\quad Q_{s_1}=\mathscr{N}_IQ^I,\quad Q_{s_2}=\mathscr{B}_IQ^I,
	\end{equation} 
	where the Mukhanov-Sasaki variables are defined as $Q^I=\delta\phi^I+\left(\dot{\phi}^I/H\right)\Phi$ with $I=1,2,3$ labeling the scalar fields.
	Finally, by rescaling the perturbations in (\ref{eq:gaugeperturbations}), one can find the curvature, first entropy and second entropy perturbations
	\begin{equation}
		(\mathcal{R},\mathcal{S}_1,\mathcal{S}_2)\equiv\frac{1}{\sqrt{2\epsilon}M_P}(Q_\sigma,Q_{s_1},Q_{s_2}).
	\end{equation}
	Although the entropy modes $(\mathcal{S}_1,\mathcal{S}_2)$ are not physical observables, their power spectra, similar to the curvature power spectrum at the end of inflation, are given by 
	\begin{align}
		\mathcal{P}_\mathcal{R}(k)&=\frac{k^3}{2\pi^2}|\mathcal{R}_k(t)|^2\bigg|_{t=t_\mathrm{end}},\\
		\mathcal{P}_{\mathcal{S}_1}(k)&=\frac{k^3}{2\pi^2}|{\mathcal{S}_1}_k(t)|^2\bigg|_{t=t_\mathrm{end}},\\
		\mathcal{P}_{\mathcal{S}_2}(k)&=\frac{k^3}{2\pi^2}|{\mathcal{S}_2}_k(t)|^2\bigg|_{t=t_\mathrm{end}}.
	\end{align}
	
	In order to reveal the hierarchical structure induced by the Frenet–Serret system in the interactions of curvature and entropy perturbations, we write the second-order action. For this purpose, we consider the following decomposition of field perturbations
	\begin{align}
		\delta\phi^a(t,\mathbf{x})=\mathscr{T}^a\wp(t,\mathbf{x})+\mathscr{N}^a\theta(t,\mathbf{x})+\mathscr{B}^a\Lambda(t,\mathbf{x}).
	\end{align}
	In the comoving gauge, field perturbations totally belong to the entropy sector, i.e. $\wp=0$. In this gauge, we write the Arnowitt-Deser-Misner (ADM) metric in terms of the curvature perturbation $\zeta$
	\begin{align}
		ds^2=-N^2dt^2+a^2e^{2\zeta}\delta_{ij}\left(dx^i+N^idt\right)\left(dx^j+N^jdt\right),
	\end{align}
	where $N$ and $N^i$ are the lapse function and shift vector, respectively \cite{Arnowitt:1962hi}. Solving the constraint equations, one can obtain the second-order Lagrangian of the curvature perturbation $\zeta$ and entropy modes $\mathscr{S}^\alpha=(\theta,\Lambda)$,
	\begin{align}\label{eq:O2Lagrangian}
		\mathcal{L}^{(2)}=a^3\left[\epsilon M^2_P\left(\dot{\zeta}^2-\frac{\left(\nabla\zeta\right)^2}{a^2}\right)+\frac{1}{2}\left(\dot{\theta}^2+\dot{\Lambda}^2-\frac{(\nabla\theta)^2}{a^2}-\frac{(\nabla\Lambda)^2}{a^2}\right)+2\sqrt{2\epsilon}\Omega\theta\dot{\zeta}M_P+\Xi\left(\theta\dot{\Lambda}-\dot{\theta}\Lambda\right)-\frac{1}{2}m^2_{\alpha\beta}\mathscr{S}^\alpha\mathscr{S}^\beta\right],
	\end{align}
	with the entropy mass matrix given by
	\begin{align}\label{eq:EnropyMass}
		m^2_{\alpha\beta}\equiv
		\begin{pmatrix}
			V_{cd}\mathscr{N}^c\mathscr{N}^d-\Omega^2-\Xi^2-2\epsilon H^2M^2_P\;\mathbb{R}_\mathscr{TNTN}&V_{cd}\mathscr{N}^c\mathscr{B}^d-2\epsilon H^2M^2_P\;\mathbb{R}_\mathscr{TNTB}\\
			V_{cd}\mathscr{N}^c\mathscr{B}^d-2\epsilon H^2M^2_P\;\mathbb{R}_\mathscr{TNTB}&V_{cd}\mathscr{B}^c\mathscr{B}^d-\Xi^2-2\epsilon H^2M^2_P\;\mathbb{R}_\mathscr{TBTB}
		\end{pmatrix},
	\end{align}
	where $\mathbb{R}_{WXYZ}\equiv\mathbb{R}_{abcd}W^aX^bY^cZ^d$ is the field space Riemann tensor contracted with the Frenet-Serret basis vectors. As is clear in (\ref{eq:O2Lagrangian}), the first entropy mode is the only portal between the curvature and entropy sector when the background trajectory undergoes turns ($\Omega\neq0$)\footnote{
	Using (\ref{eq:zeta}), one can show that in the superhorizon limit ($k\ll aH$): $\dot{\zeta}\simeq-\sqrt{\frac{2}{\epsilon M^2_P}}\Omega\;\theta$.
	}.
	In fact, this result is a consequence of the Frenet-Serret system and can be obtained even in the inflationary models with more than three fields \cite{Pinol:2020kvw}. However, at the cubic level the second entropy ($\Lambda$ or equivalently $Q_{s_2}$) can interact with the curvature perturbation. The equations of motion of the perturbations $(\zeta,\theta,\Lambda)$, obtained from the second-order action $S^{(2)}=\int dtd^3x\mathcal{L}^{(2)}$, are given by 
	\begin{align}
		&\ddot{\zeta}+(3+\eta)H\dot{\zeta}+\frac{k^2}{a^2}\zeta=-\frac{1}{a^3\epsilon M_P}\left[\sqrt{2\epsilon}a^3\Omega\theta\right]^{\displaystyle{\cdot}},\label{eq:zeta}\\
		&\ddot{\theta}+3H\dot{\theta}+\left(\frac{k^2}{a^2}+m^2_{\theta\theta}\right)\theta=2\sqrt{2\epsilon}M_P\Omega\dot{\zeta}+\left(3H\Xi+\dot{\Xi}-m^2_{\theta\Lambda}\right)\Lambda+2\Xi\dot{\Lambda},\\
		&\ddot{\Lambda}+3H\dot{\Lambda}+\left(\frac{k^2}{a^2}+m^2_{\Lambda\Lambda}\right)\Lambda=-\left(3H\Xi+\dot{\Xi}+m^2_{\Lambda\theta}\right)\theta-2\Xi\dot{\theta},
	\end{align}
	where $\eta\equiv\dot{\epsilon}/(H\epsilon)$ is the second slow-roll parameter. 
	In the superhorizon regime, even in the presence of entropy sourcing, the entropy modes evolve as a closed system  
	\begin{align}
		\frac{d^2}{dt^2}
		\begin{pmatrix}
			\theta\\
			\Lambda
		\end{pmatrix} +
		\begin{pmatrix}
			3H&-2\Xi\\
			2\Xi&3H
		\end{pmatrix}
		\frac{d}{dt}
		\begin{pmatrix}
			\theta\\
			\Lambda
		\end{pmatrix}
		+
		\begin{pmatrix}
			m^2_{\theta\theta}+4\Omega^2&-3H\Xi-\dot{\Xi}+m^2_{\theta\Lambda}\\
			3H\Xi+\dot{\Xi}+m^2_{\Lambda\theta}&m^2_{\Lambda\Lambda}
		\end{pmatrix}
		\begin{pmatrix}
			\theta\\
			\Lambda
		\end{pmatrix}\simeq0.
	\end{align}
	\section{The model}
	
	\begin{table}
		\centering
		\caption{Parameters used in the model (\ref{eq:two-peak}). All the masses and parameters are given in units where $M_P=1$.}
		\label{tab:params}
		\renewcommand{\arraystretch}{1.5}
	\begin{tabular}{| c | c | c | c | c | c | c | c | c |}
	\hline
	Parameter   & $m_\chi$ &  $m_\psi$ & $V_0$ & $\phi_0$ &$\mathrm{g}^2$ &$\beta$ &$\upsilon$ &$\alpha$\\
	\hline 
	Value & $2.9916\times10^{-6}$ & $1.5991\times10^{-7}$ &  $3.58\times10^{-9}$&$\sqrt{6}$ &$8\times10^{-13}$ &7.685&6.850&6.868\\
	\hline
\end{tabular}
\end{table}
	In order to have two localized features in the evolution of the background and consequently a double-peaked power spectrum, one can consider a three-phase dynamics. This aim can be achieved within the context of three-field inflation. Inflation in each phase is driven by only one of the scalar fields, while the others remain frozen. If a mass hierarchy (with a mass ratio of approximately greater than 7) is assumed for the fields, the background dynamics passes through transient epochs at the times of phase transitions. As the effective mass of isocurvature modes in (\ref{eq:EnropyMass}) shows, if a suitable combination of field space curvature and potential derivatives is present, isocurvature modes can play an important role in the dynamics of perturbations. Therefore, by setting a linear dependence in the field space metric with $b(\phi)\equiv\beta\phi$, $y(\phi)\equiv\upsilon\phi$ and $p(\chi)\equiv\alpha\chi$, we consider this Lagrangian
	\begin{equation}\label{eq:two-peak}
		\mathcal{L}_\phi=-\frac{1}{2}\partial_\mu\phi\partial^\mu\phi-\frac{1}{2}e^{2\beta\phi}\partial_\mu\chi\partial^\mu\chi-\frac{1}{2}e^{2\upsilon\phi+2\alpha\chi}\partial_\mu\psi\partial^\mu\psi-V_0\frac{\phi^2}{\phi^2+\phi_0^2}-\frac{1}{2}m^2_\chi\chi^2-\frac{1}{2}m^2_\psi\psi^2-\frac{1}{2}\mathrm{g}^2\phi^2\psi^2,
	\end{equation}
	which can be viewed as a simple generalization of the model studied in \cite{Braglia:2020eai}.
	The parameters used in this model (summarized in Table \ref{tab:params}) are tuned for the discussed purpose. The scale factor $a(t)$ is normalized such that the pivot scale $k_\ast=0.05\;\mathrm{Mpc}^{-1}$ exits the horizon $50$ e-folds before the end of inflation. 
	The codes utilised in this study are available in \cite{Nasrolahzadeh:2026code}.
	\begin{figure}[t]
		\includegraphics[width = 0.85\textwidth]{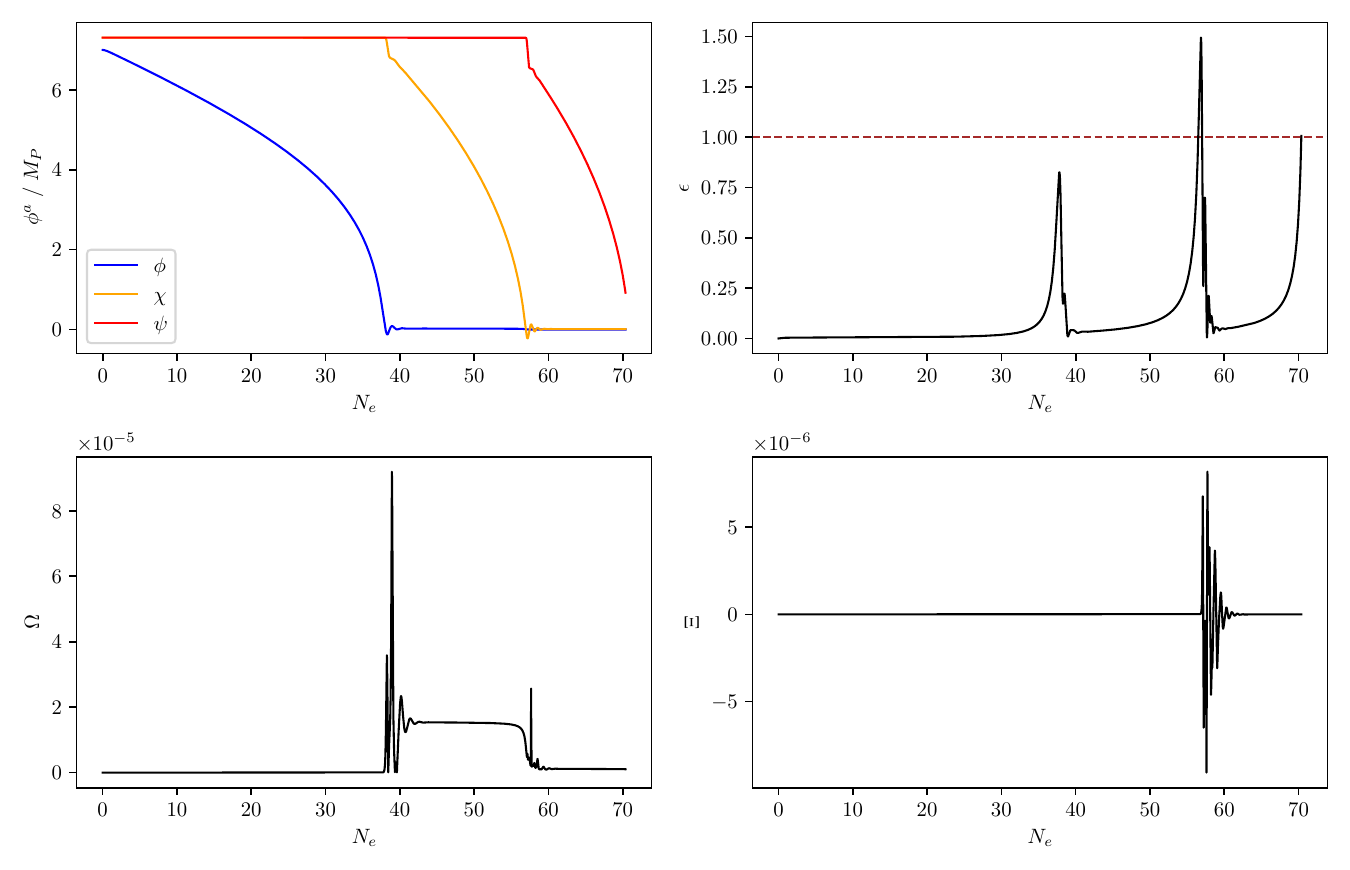}
		\centering
		\caption{Temporal evolution of background quantities, the scalar fields $\phi^a$ (\textit{top left}), slow-roll parameter $\epsilon$ (\textit{top right}), turning rate $\Omega$ (\textit{bottom left}) and torsion $\Xi$ (\textit{bottom right}). The background trajectory starts from the initial point $\phi^a_\mathrm{i.c}=(7,7.31,7.31)M_P$ in the field space with zero velocity and contains three phases, each phase due to rolling-down of one field, and in the other view, three stages of accelerated, decelerated and again accelerated expansion.}
		\label{fig:Background}
	\end{figure}
	\begin{figure}
		\includegraphics[width = 0.4\textwidth]{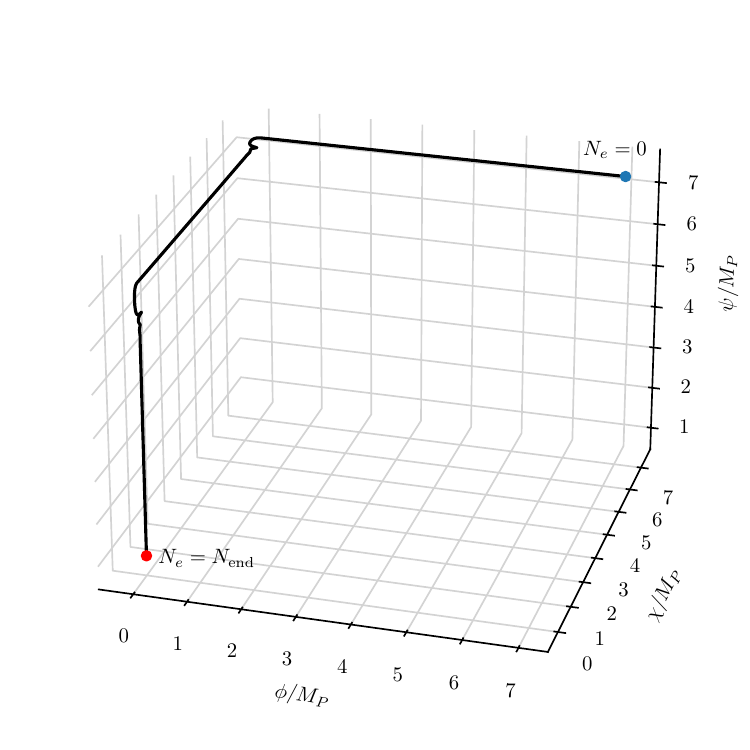}
		\centering
		\caption{The background trajectory in 3D field space. Note that the field space due to non-trivial metric $\mathcal{G}$ is not flat, despite the Euclidean appearance of this plot.}
		\label{fig:3DBackground}
	\end{figure}
	\begin{figure}[t]
		\includegraphics[width = 0.9\textwidth]{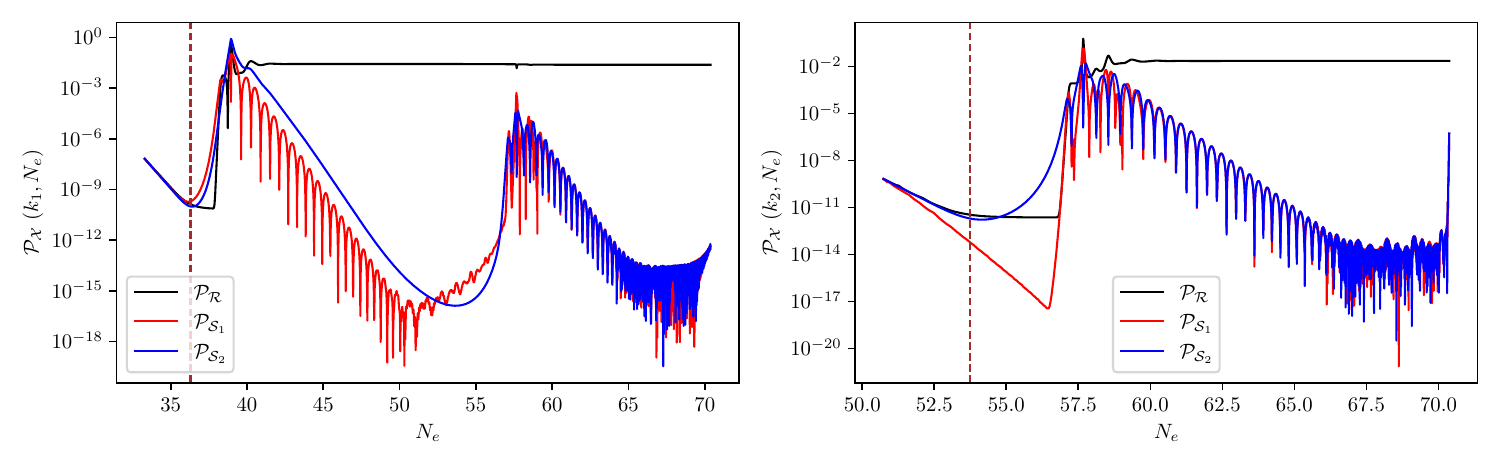}
		\centering
		\caption{Time evolution of the curvature and isocurvature power spectra for two specific modes of $k_1=2.6\times10^{5}\mathrm{Mpc}^{-1}$ (\textit{left}) and $k_2=1.4\times10^{12}\mathrm{Mpc}^{-1}$ (\textit{right}). The black, red and blue lines indicate evolution of power spectrum of curvature, first entropy and second entropy, respectively. The dashed line shows the time of horizon exiting ($k=aH$) in e-fold.}
		\label{fig:PertTrack}
	\end{figure}  
	\begin{figure}[t]
		\includegraphics[width = 0.65\textwidth]{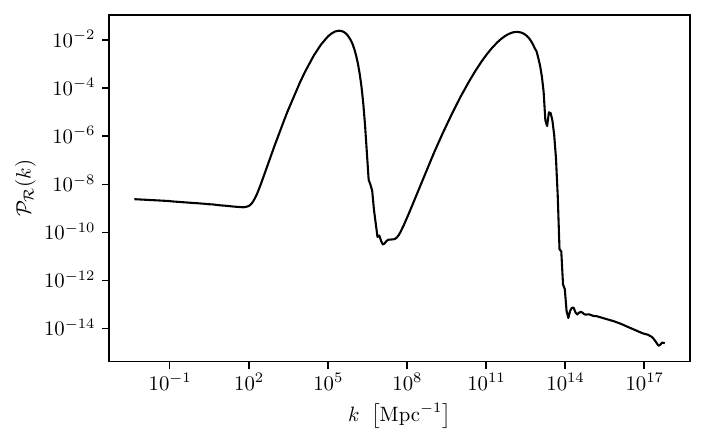}
		\centering
		\caption{The curvature power spectrum at the end of inflation.}
		\label{fig:3F_power2peaks}
	\end{figure}
	
	The temporal evolution of background quantities is shown in FIG. \ref{fig:Background}. The first field $\phi$ rolls to the minimum of its potential, while the others are frozen until $\phi$ reaches the minimum and $\chi$ starts rolling down, and so on. At the first phase transition, the background passes through a transient epoch in which the slow-roll parameter undergoes a jump and the background trajectory turns into the $\chi$-direction. But at the second phase transition around $N_e\sim57$, when the condition $\epsilon<1$ breaks, inflation ends and a brief stage of decelerated expansion starts. This time the background trajectory, with a smaller value of turning rate but a large torsion gets aligned with the $\psi$-direction. The coincident spikes in turning rate and torsion in the second epoch do not have the top hat shape, despite the case studied in \cite{Aragam:2023adu,Aragam:2024nej}. Indeed, since the field $\psi$ has no significant dynamics in the first two phases, the background trajectory remains in the two dimensional $\phi-\chi$ subspace at the first phase transition without large torsion (see FIG. \ref{fig:3DBackground}).
	
	As expected, two transient epochs in the evolution of the background affect the evolution of perturbations (see FIG. \ref{fig:PertTrack}). To see this effect, we consider two modes $k_1=2.6\times10^{5}\mathrm{Mpc}^{-1}$ and $k_2=1.4\times10^{12}\mathrm{Mpc}^{-1}$ which leave the Hubble radius before each epoch, respectively. Once the mode $k_1$ leaves the horizon, it enters the first epoch and all of the curvature and isocurvature perturbations get amplified and by exiting this era, isocurvature modes damp and curvature freezes until the mode reaches the second epoch. Although the isocurvature modes get amplified again, they cannot influence the curvature much this time. The mode $k_2$ just feels the second transient epoch. When the mode reaches this era, all the perturbations get amplified, and then the isocurvature ones decay (note that the sudden growth at the end of inflation has no physical cause and happens due to numerical instability) and curvature freezes at the order of $\mathcal{O}(0.02)$. 
	These two modes are representative of two different classes of modes that see these epochs, and curvature at those scales gets amplified. Roughly speaking, the role of turns in these amplifications is to rearrange the type of field perturbations $(\delta\phi,\delta\chi,\delta\psi)$ in an adiabatic or entropy manner. For example, $\delta\phi$ is initially adiabatic, but after the first turn it becomes entropy, whereas the initially entropic $\delta\chi$ becomes adiabatic.   
	
	The overall curvature power spectrum is shown in FIG. \ref{fig:3F_power2peaks}. It exhibits two distinct peaks with amplitudes appropriate for PBH formation.
	\section{PBH formation}
	A large enough perturbative mode of $k$ can undergo its own gravitational collapse when it re-enters the horizon ($k=aH$) and form a PBH. The associated mass with that PBH can be expressed as
	\begin{align}
		M(k)=30\left(\frac{\gamma}{0.2}\right)\left(\frac{g_{\ast,f}}{10.75}\right)^{-1/6}\left(\frac{k}{2.9\times10^5\;\mathrm{Mpc}^{-1}}\right)^{-2}M_\odot,
	\end{align}
	where $\gamma=0.2$ is the correction factor for the fraction of PBH mass in the horizon mass, and $g_{\ast,f}$ is the number of relativistic degrees of freedom at the time of formation \cite{Sasaki:2018dmp}. According to the Press-Schechter formalism, the fraction of PBHs in energy density for a Gaussian profile can be written as 
	\begin{align}
		\beta(M)=2\gamma\int_{\delta_c}\frac{e^{-\frac{\delta^2}{2\sigma^2}}}{\sqrt{2\pi}\sigma}d\delta,
	\end{align}
	where $\delta_c$ is the threshold of PBH formation and $\sigma$ here is the variance of perturbations
	\begin{align}
		\sigma^2(k)=\frac{16}{81}\int d\ln qW^2(q/k)(q/k)^4\mathcal{P}_\mathcal{R}(q),
	\end{align}
	where $W$ is the window function. Finally, the fraction of PBHs in dark matter (DM) is given by
	\begin{align}
		f_{\mathrm{PBH}}(M)\equiv\frac{\Omega_{\mathrm{PBH}}}{\Omega_\mathrm{DM}}=2.7\times10^8\left(\frac{\gamma}{0.2}\right)^{1/2}\left(\frac{g_{\ast,f}}{10.75}\right)^{-1/4}\left(\frac{M}{M_\odot}\right)^{-1/2}\beta(M).
	\end{align}
	\begin{figure}[t]
		\includegraphics[width = 0.55\textwidth]{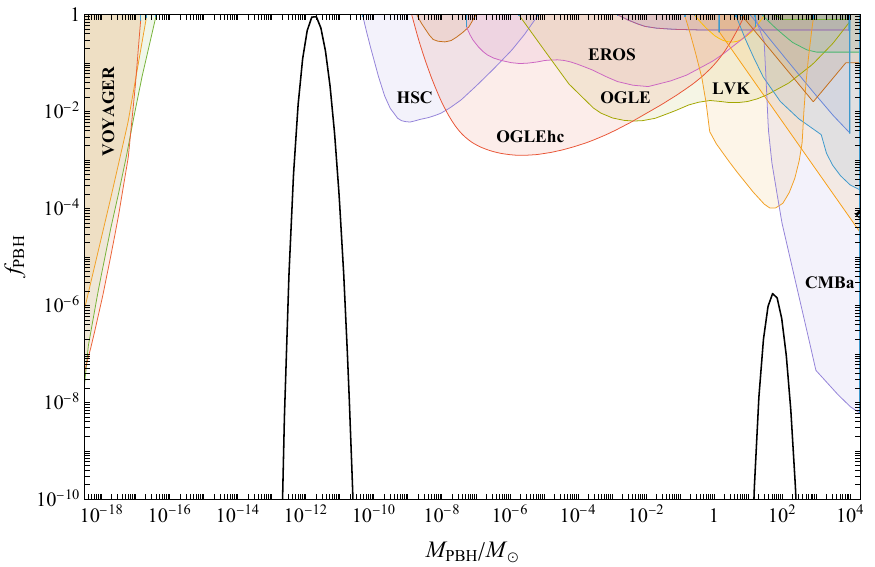}
		\centering
		\caption{Fraction of PBHs in dark matter. The observational constraints are taken from \cite{Carr:2026hot}. The solid line shows a bimodal mass spectrum for $f_\mathrm{PBH}$ in two distinct mass ranges. The left peak lies completely within the open window that has no observational constraints yet, constituting a large fraction of dark matter. The right peak lies in the mass range consistent with the black hole binary masses detected in GW observations.}
		\label{fig:constraints}
	\end{figure}
	
	The abundance of PBHs resulting from the power spectrum in FIG. \ref{fig:3F_power2peaks} is shown in FIG. \ref{fig:constraints}. The plot shows that the double-peak curvature power spectrum yields a mass spectrum with two characteristic masses $M_1\sim2\times10^{-12}M_\odot$ and $M_2\sim52M_\odot$, explaining simultaneously both DM and GWs, respectively\footnote{A non-inflationary break, in principle, can yield the formation of another family of ultralight PBHs through a twice horizon re-entry mechanism \cite{Wang:2025lti}. However the duration of intermediate stage in our work is not sufficiently long for that purpose and hence we do not expect a third population of PBHs to be formed.}. The total fraction of PBHs from these two mass windows is $f^\mathrm{tot}_\mathrm{PBH}=\int d\ln Mf_\mathrm{PBH}(M)\simeq0.98$. The amplitude of $f_\mathrm{PBH}$ around $M_2$ can be larger by choosing another point in the parameter space (e.g. with higher $\upsilon$), but this may yield a brief breakdown of linear theory in the evolution of isocurvature modes during the first turn.
	\section{Discussion and Conclusion}
	In this work, we have studied a three-field framework for inflation in a curved field space. By using the kinematic (Frenet-Serret) system, we characterized the background trajectory in three-dimensional field space through its two geometric quantities: turning rate and torsion. We derived the equations of motion for curvature and isocurvature perturbations within this system. We explicitly highlighted the hierarchical structure among the interactions of perturbations, as it exists in the mathematical structure of the kinematic basis: the second entropy mode does not couple to the curvature perturbations directly and instead influences it through sourcing the first entropy mode.
	This finding may suggest that even in the multifield models with $\mathcal{N}\geq4$, the physics of curvature perturbations can be effectively explained by the curvature-first entropy interaction.
	
	
	
	We investigated a three-phase model, with each phase driven by only one field while the other two fields remain frozen, producing a three-stage dynamics of accelerated, decelerated and accelerated expansion during inflation. We showed that under a suitable choice of interactions, two families of perturbation modes get amplified during the phase transitions, leading to a double-peaked curvature power spectrum with $\mathcal{P}_\mathcal{R}\sim\mathcal{O}(0.02)$ at both peaks. These enhancements are due to power transfer from isocurvature modes to the curvature mode, i.e. mode conversion. The consequence of double-peaked behavior in the power spectrum is the formation of two families of PBHs ($M_1\sim2\times10^{-12}M_\odot$ and $M_2\sim52M_\odot$), explaining DM and GWs, simultaneously, with the same origin. As we mentioned earlier, our results depend on the chosen parameters. It is fine-tuned but shows the possibility of having both DM and GW results. However, it is interesting to see if this result can be a generic prediction for a class of multifield inflationary scenarios.
	
	
	In this paper, we did not compute the amount of non-Gaussianity produced in the presented model. Because of the importance of perturbations profile in the abundance of PBHs \cite{Young:2013oia,Yoo:2019pma}, it would be interesting to investigate the three-point correlation function (bispectrum) in future works. 
	\section*{Acknowledgment} \label{sec:ack}                                            
	We would like to thank Antonio Riotto and Misao Sasaki for their comments on this manuscript.
	\appendix

\end{document}